\documentclass[sigconf, table, dvipsnames]{acmart}
\usepackage{acro}
\usepackage{booktabs}
\usepackage{graphicx}
\usepackage{balance}
\usepackage{titlesec}
\usepackage{hyperref}
\usepackage{multirow}
\usepackage{siunitx}
\usepackage{subcaption}

\usepackage{graphicx}
\usepackage{svg}
\usepackage{booktabs}
\usepackage{textcomp}
\usepackage{algorithm}
\usepackage[noend]{algpseudocode}
\usepackage{float} 
\usepackage{array}
\usepackage{soul}
\usepackage{tikz}

\usepackage{graphicx}
\AtBeginDocument{%
  }

\copyrightyear{2025}
\acmYear{2025}
\setcopyright{cc}
\setcctype{by-nc-sa}
\acmConference[WWW Companion '25]{Companion Proceedings of the ACM Web
Conference 2025}{April 28-May 2, 2025}{Sydney, NSW, Australia}
\acmBooktitle{Companion Proceedings of the ACM Web Conference 2025 (WWW
Companion '25), April 28-May 2, 2025, Sydney, NSW, Australia}
\acmDOI{10.1145/3701716.3717858}
\acmISBN{979-8-4007-1331-6/2025/04}

\begin{document}



\title{Middleman Bias in Advertising: Aligning Relevance of Keyphrase Recommendations with  Search\\
}

\author{Soumik Dey}
\affiliation{%
  \institution{eBay Inc.}
  \city{San Jose}
  \state{CA}
  \country{USA}
}
\email{sodey@ebay.com}

\author{Wei Zhang}
\affiliation{%
  \institution{eBay Inc.}
  \city{Shanghai}
  \country{China}}
\email{wzhang15@ebay.com}

\author{Hansi Wu}
\affiliation{%
  \institution{eBay Inc.}
  \city{San Jose}
  \state{CA}
  \country{USA}
}
\email{hanswu@ebay.com}

\author{Bingfeng Dong}
\affiliation{%
 \institution{eBay Inc.}
  \city{Shanghai}
  \country{China}}
\email{bidong@ebay.com}

\author{Binbin Li}
\affiliation{%
  \institution{eBay Inc.}
  \city{San Jose}
  \state{CA}
  \country{USA}}
\email{binbli@ebay.com}

\begin{abstract}
E-commerce sellers are recommended keyphrases based on their
inventory on which they advertise to increase buyer engagement
(clicks/sales). Keyphrases must be pertinent to items; otherwise, it
can result in seller dissatisfaction and poor targeting — towards
that end relevance filters are employed. In this work, we describe
the shortcomings of training relevance filter models on biased
click/sales signals. We re-conceptualize advertiser keyphrase relevance as interaction between two dynamical systems — Advertising
which produces the keyphrases and Search which acts as a middleman to reach buyers. We discuss the bias of search relevance systems (middleman bias) and the need to align advertiser keyphrases
with search relevance signals. We also compare the performance of
cross encoders and bi-encoders in modeling this alignment and the
scalability of such a solution for sellers at eBay.
\end{abstract}



\maketitle

\section{Introduction}
\label{sec:Introduction}

\begin{figure*}[t]
\centering
\begin{subfigure}{8.5cm}
\centering
\includegraphics[width=\linewidth]{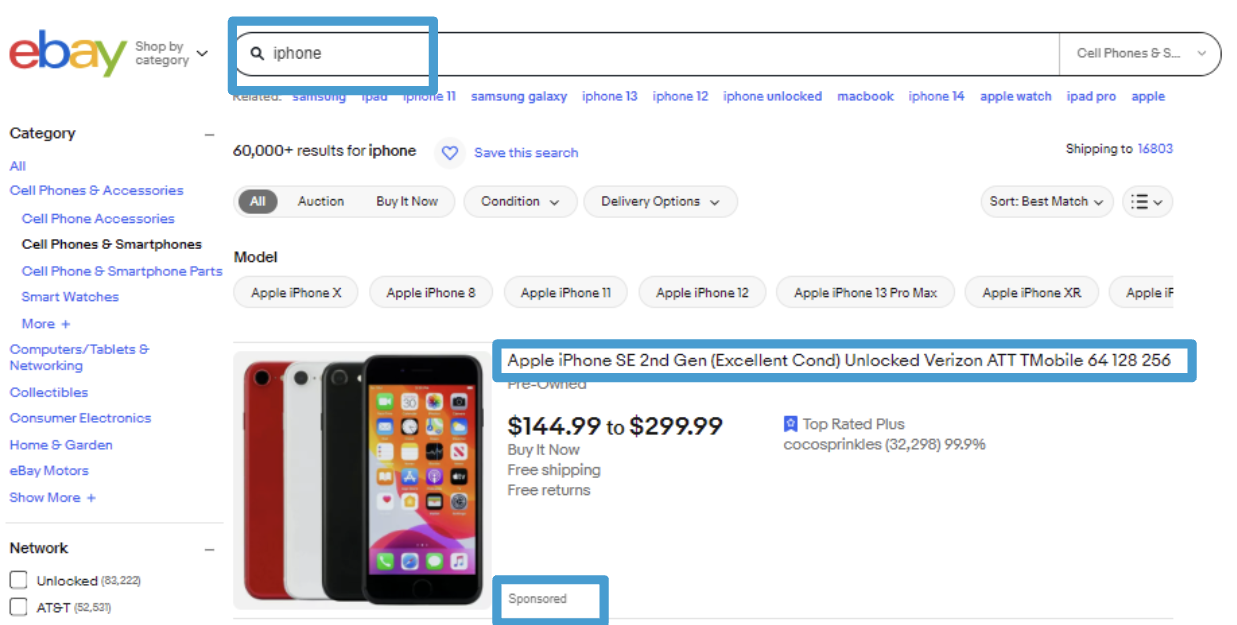}
  \caption{Buyer side}
  \label{fig:1a}
\end{subfigure}\qquad
\begin{subfigure}{8.5cm}
\centering

\includegraphics[width=\linewidth]{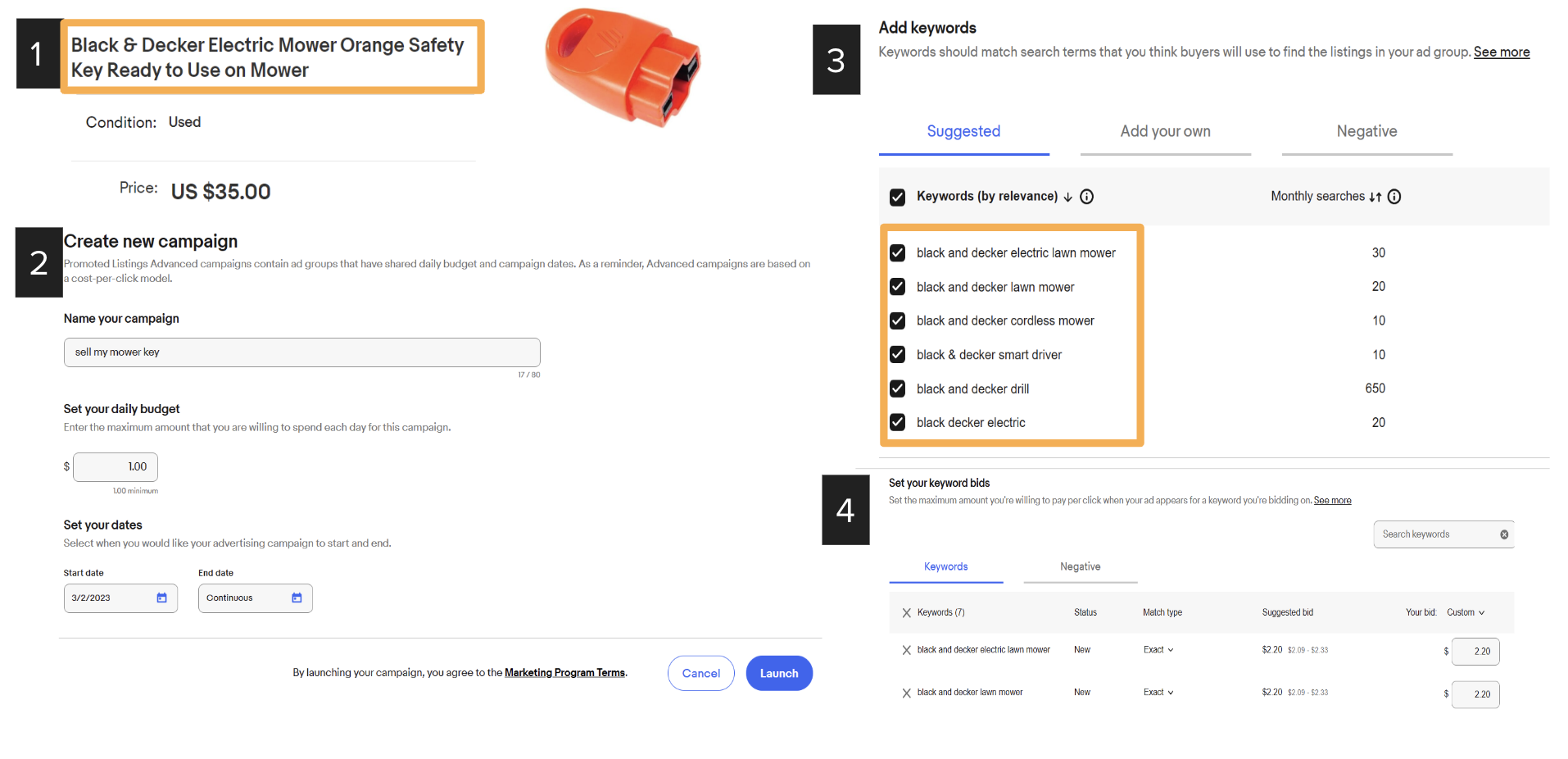}
\caption{Seller Side}
\label{fig:1b}
\end{subfigure}
\caption{Screenshot of our keyphrases for manual targeting in Promoted Listings Priority\textsuperscript{TM} \cite{eBay} for eBay Advertising.}
\label{fig:screenshot}
\end{figure*}

Within the landscape of e-commerce, sellers utilize online advertising mechanisms such as keyphrase recommendations to overcome low organic search placement, thereby gaining advantageous visibility on the search results page (SRP) and enhancing buyer interaction. Advertiser Keyphrase recommendation at eBay involves three primary steps: \textit{retrieval}, which collects keyphrases relevant to the item; \textit{relevance}, which further filters these keyphrases to ensure their pertinence to the item, and \textit{ranking}, which encourages preferential seller adoption based on the seller’s budget and preferences. The significance of keyphrase relevance is pivotal, as it influences seller perception and helps prevent Search systems from being inundated with numerous non-relevant items competing in auctions.

Advertiser keyphrase relevance models can be trained to identify general trends in click/sales data. A keyphrase that has garnered a sufficient number of clicks or sales in relation to an item can be considered relevant to the item. In essence, clicks and sales are strong positive indicators of relevance. However, they do not serve well as indicators of irrelevance. E-Commerce data suffer from missing-not-at-random (MNAR) conditions due to several biases \cite{popularitybias, zhang2024lazyprolifictacklingmissing, bias_survey}. An item lacking clicks for a particular query does not necessarily mean that it is irrelevant. In the context of e-Commerce, buyers act as annotators, but unlike typical annotators, they see a skewed presentation of items influenced by search ranking, which in turn affects their annotations (clicks/sales). An unpopular item will have a poor position on the Search results page (SRP) due to rankings being based on clicks and subsequently may not obtain any clicks or sales. This bias results in the unreliability of negative relevance signals based on clicks or sales. 

\begin{figure}[b]
\centering
\includegraphics[width=\linewidth]{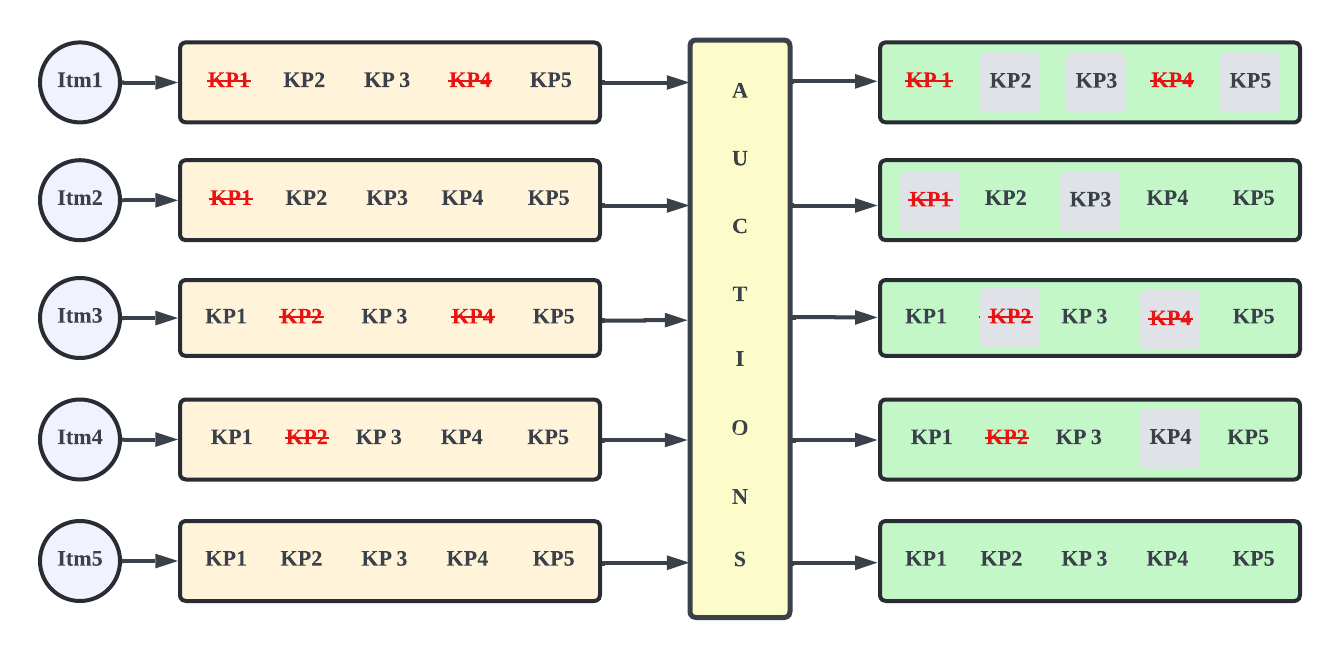}
\caption{Auctions of items (Itm) in relation to keyphrases (KP). Red strikethrough font
represents filter of Advertising while gray highlight represents that of Search. Both gray highlight and red-strikethrough represents kephrases eliminated by both Advertising and Search filters.}
\label{fig:auctions}
\end{figure}  

 \textit{eBay Search} shows items in response to buyer queries and also involves the same tasks: \textit{retrieval}, retrieving items related to the search queries; \textit{relevance}, a more detailed filtering of the retrieved items to ensure conformity to the query, and \textit{ranking} , i.e. ordering the retrieved items based on potential to achieve clicks so that buyers interact with them preferentially. From the seller's perspective, when \textit{eBay Advertising} presents advertised keyphrases to sellers, the sellers choose to bid on them for their items. These items then enter an auction corresponding to the keyphrase. This auction process involves a complex interaction with \textit{eBay Search}. During this auction, \textit{eBay Search} acts as a \textit{middleman} and aligns the items with the queries, which are exactly the advertised keyphrases recommended by us, and further filters the auction items using their relevance filter. This auction mechanism also ensures that the click data logged will only contain keyphrases which pass Search's relevance filter (auction winning impression gaining keyphrases are logged). \textit{This introduces an additional bias --- training will only be on keyphrases which pass the Search relevance filter}. Training on click data ensures that the model never experiences keyphrases that are irrelevant to Search, while Advertising does generate such keyphrases and needs to filter them. We coin this bias as \textit{middleman bias} a form of sample selection bias~\cite{rec4ad, sample_selection}, which the click data additionally suffers from, in the context of advertiser keyphrase recommendation.

The interaction between Advertising and Search displays an inherent imbalance, as Search ultimately determines which keyphrases are deemed relevant. For instance, take \texttt{Itm1} from Figure \ref{fig:auctions}; Advertising considers \texttt{KP1} and \texttt{KP4} irrelevant, while \textit{eBay Search} regards \texttt{KP2}, \texttt{KP3}, and \texttt{KP5} as irrelevant, leading to \texttt{Itm1} not participating in any auctions. \textit{This outcome occurs regardless of the true relevance of the keyphrases, meaning that even if \texttt{KP2}, \texttt{KP3}, and \texttt{KP5} are genuinely relevant, \texttt{Itm1} will not enter those auctions because Search does not find them relevant}. Therefore, it is prudent to align the relevance filter of Advertising with that of Search, which acts as a \textit{middleman} in this context. In scenarios with optimal outcomes, such as \texttt{Itm3} and \texttt{Itm5}, there is absolute agreement between Advertising and Search. As discussed, Advertising requires debiasing the click-based performance data, while still filtering out keyphrases deemed irrelevant by Search. We thus redefine the relevance of keyphrase recommendations as a complex interaction between Advertising and Search and propose a scalable solution to align these systems --- reminiscent of alignment of dependent systems in cascade ranking~\cite{cascade}.

\section{Previous Experiments and Learnings}

In this section, we discuss some previous experiments and their shortcomings and describe our learnings from them. A previous BERT model was trained on human judgment (about 80,000 records with 3 annotators). However, after the launch of the model, we received multiple complaints from sellers about irrelevant keyphrases. We had to pull the model from production and a post launch analysis revealed that due to the large number of categories at eBay; some categories had inadequate representation in this dataset of only 80,000 records. In addition, the data suffered from low agreement between the annotators, making it difficult to train on the data.\footnote{The diversity of eBay items across more than 100,000 categories makes it necessary to collect a lot more than 80,000 records to ensure there is proper representation.}

We also experimented with an xgboost-based CTR model trained on click data. Offline evaluation of the CTR model failed in terms of search alignment and spot-checked human judgement. An alternate model based on the Jaccard index \cite{jaccard1901etude} proved better and was released to production. The Jaccard index was a simple rule-driven token-based algorithm with a threshold to filter out irrelevant keyphrases. This worked, since token similarity is an important feature for relevance. Notably, the CTR model had Jaccard index as a feature but still underperformed. Further analysis indicated that the CTR model was significantly biased, lacking representation from item-query pairs that failed the search relevance filter (required to stay in the auction), and carried forward the described e-Commerce data biases described in Section \ref{sec:Introduction}.

%We conducted experiments using a CTR model based on xgboost, tied to biased click data, as outlined earlier.ootnote{Details of the model are beyond the scope of this work.} During offline evaluation, the CTR model failed in search alignment and human judgement checks. An alternative model utilizing the Jaccard filter outperformed it and was deployed in production. The Jaccard index, a simple token-based algorithm driven by rules and a threshold, effectively filtered out irrelevant keyphrases due to the importance of token similarity for relevance. Notably, the CTR model included the Jaccard index as a feature but still underperformed. Further analysis indicated that the CTR model was significantly biased, lacking representation from item-query pairs that failed the search relevance filter (required to enter the auction) and carried forward the described e-Commerce data biases.

Jaccard index defined as the token intersection over union has its own problems. Since item titles are generally longer than keyphrases, it heavily penalizes shorter keyphrases --- which are generally head keyphrases (head keyphrases are keyphrases which drive most clicks/sales, at eBay 10\% of keyphrases drive 90\% of buyer engagement in clicks/sales). The introduction of the jaccard filter essentially eliminated single-token keyphrases, which were a major source of advertisement for sellers. Additionally, jaccard not being semantically aware; would miss targeting opportunities on those keyphrases semantically aligned with the item. Jaccard is easy to implement and integrate in eBay systems; however, it is not the most accurate relevance model and had been adopted as a stop gap.

\section{Data Curation and Experiment Design}
Reminiscent of other methods of debiasing by curating data from additional sources \cite{rec4ad}, to circumvent all the problems described in Section \ref{sec:Introduction} we use a dataset of Search's relevance judgment on our Advertisement recommendations to mitigate \textit{middleman bias} in Advertising. This ensures that our model does not suffer from the same biases as the click data, while at the same time ensuring that we align with Search's relevance judgment for the most optimal outcomes in auctions for our keyphrase recommendations. We curated a dataset of 24 million and 3 million item-keyphrase pairs with labels 1 and 0 indicating the pass / fail of Search's relevance judgment for training and evaluation, respectively. The data is stratified by categories based on their activity in terms of traffic. This huge data guarantees proper representation of all categories while still maintaining preferential bias of models towards high performing categories \cite{ashirbad-etal-2024}. The problem we are trying to solve for is --- given an item title, its category information, and the keyphrase associated, can we predict whether the keyphrase would be deemed relevant/irrelevant by Search?

For our item data, we take into account the \texttt{title} of the item and the \texttt{category name}. Previous work in \cite{ashirbad-etal-2024, mishra2024graphexgraphbasedextractionmethod} has shown how categorical information enriches item data. In terms of representing textual information and modeling for a classification (NLU/NLI); encoders have been the popular choice. The two most popular frameworks for generating text representations using encoders are: 1) \textit{bi-encoders}, which separately computes the embeddings of item ($u$) and the keyphrase ($v$) with the input of the item encoding being \texttt{title [SEP] category name}, and 2) \textit{cross-encoders}, where embedding of item and keyphrase are processed together, the input is \texttt{keyphrase [SEP] category name [SEP] title}. 

In employing bi-encoders as a base model, we utilized the eBay pre-trained BERT model, \textit{eBERT} \cite{dahlmann2021deployingbertbasedquerytitlerelevance} and fine-tuned using:

\begin{itemize}
    \item \textit{Contrastive learning}\cite{contrastive} on search relevance scores of item-keyphrase pairs which minimizes the distance between the embeddings with label 1 and maximizes the distance between the embeddings with label 0.
    \item \textit{Softmax Classifier} trained on concatenation and difference of item embeddings ($u$) and keyphrase embeddings ($v$), i.e. $(u, v, |u-v|)$. This has been proven to work well for NLU tasks\cite{reimers-gurevych-2019-sentence}.
    \item \textit{In-batch Random Negative Sampling (IRNS)} based on click data \cite{inbatchnegatives}. Since click data is biased for negative labels, in-batch-negative sampling paves the way for negative construction in an unsupervised manner from all other data points in the batch. This has been shown to often yield better results \cite{reimers-gurevych-2020-making}. Towards that end we additionally curated a dataset of 24 million records from 30 days of search activity data with at least one click and with a lower lower bound of 0.05 CTR and a minimum of 30 impressions for click-based IRNS training. This is to ensure that we eliminate noisy data points, such as 1 impression, 1 click (accident), or 1000 impressions, 1 click (low CTR), which are misleading. The resulting data points can be construed as positives accumulated from search logs, which will then be used for in-batch negatives sampling.  
\end{itemize}

Since cross-encoders are expensive in terms of throughput, we experiment with cross-encoders of two sizes --- \texttt{bert-mini}\footnote{https://huggingface.co/prajjwal1/bert-mini} with 4 layers and hidden size of $256$ and \texttt{bert-tiny}\footnote{https://huggingface.co/prajjwal1/bert-tiny} with 2 layers and hidden size of $128$ from \cite{bhargava2021generalization, well-read}. These models are chosen because of their lower inference latency; more details to be discussed in Section \ref{sec:results}. All models were trained on $4$ epochs with a learning rate of $2\times 10^{-5}$ with $8$ A100-80GB GPUs. The bi-encoder models were trained using a per-device batch size of $384$ and fp16 quantization. The \texttt{bert-mini} cross encoder was trained with a batch size of $10240$ while the \texttt{bert-tiny} model was trained with a batch size of $40960$. The varying batch sizes for different models are designed to maximize the GPU memory usage for each of these models.

For our experiments, we evaluated all models on our task of aligning with search relevance. Both bi-encoders and cross-encoders have their pros and cons. Bi-encoder representations are generally more scalable as the embedding generation for items and keyphrases is done independently. Cross encoders, on the other hand, can become expensive for inference as we have to perform it for item-keyphrase combinations. However, previous research \cite{reimers-gurevych-2019-sentence} has shown that cross encoders have a higher performance in terms of NLU and classification tasks, with the caveat of higher inference latency. For comparison, we also take into account the scalability of cross-encoder models and test cross-encoders of various sizes to account for their performance in relation to their viability. Given the trade-off between performance and scalability, we compare the models based on offline evaluation on Search agreement and select the model which has the best of both worlds. We then present the online A/B test results against the Jaccard model in production. 

\section{Results}
\label{sec:results}

\begin{figure*}[t]
\centering
\includegraphics[width=0.8\linewidth]{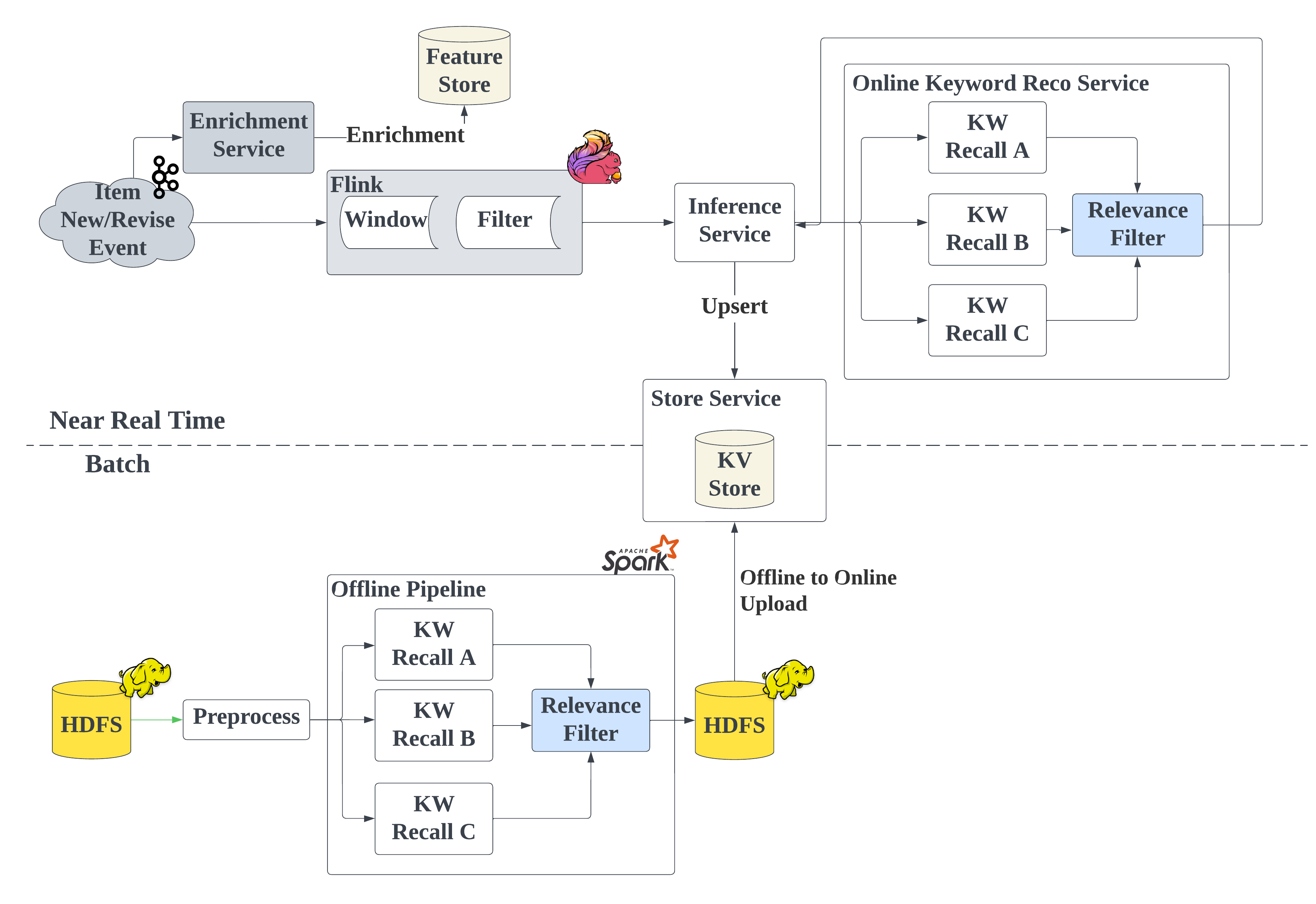}
\caption{Production Serving Architecture for our retrieval and relevance models.}
\label{fig:prod_arch}
\end{figure*}

We present the results of both bi-encoders and cross-encoders described in the previous section. We report results on an evaluation set of 3 million item-keyphrase pairs and their Search relevance judgements. The models are evaluated on precision, recall and F1 scores. While precision is the measure of how aligned Advertising's filtered keyphrases are with Search's relevance filter, recall measures the lost opportunity in terms of targeting potential of keyphrases unnecessarily filtered out. The F1 scores, which are a harmonic mean of precision and recall, are an indicator for the performance of both. 

\begin{table}[h]
\centering
\resizebox{0.47\textwidth}{!}{%
\begin{tabular}{c|c|ccccc}
\hline
 &
  Models &
  Precision &
  Recall &
  F1 &
  Diff & 
  Full \\ \cline{1-2}
 &
  Softmax &
  \cellcolor[HTML]{00D2CB}{\color[HTML]{000000} 0.60} &
  \cellcolor[HTML]{3531FF}{\color[HTML]{FFFFFF} 0.85} &
  \cellcolor[HTML]{00D2CB}{\color[HTML]{000000} 0.71} &
  \cellcolor[HTML]{000000}{\color[HTML]{FFFFFF} 1} &
  \cellcolor[HTML]{000000}{\color[HTML]{FFFFFF} 8}
  \\
 &
  IRNS &
  \cellcolor[HTML]{00D2CB}{\color[HTML]{000000} 0.62} &
  \cellcolor[HTML]{000000}{\color[HTML]{FFFFFF} 0.88} &
  \cellcolor[HTML]{3531FF}{\color[HTML]{FFFFFF} 0.73} &
  \cellcolor[HTML]{000000}{\color[HTML]{FFFFFF} 1} &
  \cellcolor[HTML]{000000}{\color[HTML]{FFFFFF} 8} \\
\multirow{-3}{*}{bi-encoder} &
  Contrastive &
  \cellcolor[HTML]{3531FF}{\color[HTML]{FFFFFF} 0.71} &
  \cellcolor[HTML]{000000}{\color[HTML]{FFFFFF} 0.88} &
  \cellcolor[HTML]{3531FF}{\color[HTML]{FFFFFF} 0.79} &
  \cellcolor[HTML]{000000}{\color[HTML]{FFFFFF} 1} & \cellcolor[HTML]{000000}{\color[HTML]{FFFFFF} 8}\\ \cline{1-2}
 &
  bert-mini &
  \cellcolor[HTML]{000000}{\color[HTML]{FFFFFF} 0.76} &
  \cellcolor[HTML]{000000}{\color[HTML]{FFFFFF} 0.88} &
  \cellcolor[HTML]{000000}{\color[HTML]{FFFFFF} 0.81} &
  \cellcolor[HTML]{00D2CB}{\color[HTML]{000000} 2.5} & 
  \cellcolor[HTML]{00D2CB}{\color[HTML]{000000} 12} \\
\multirow{-2}{*}{cross-encoder} &
  bert-tiny &
  \cellcolor[HTML]{00009B}{\color[HTML]{FFFFFF} 0.74} &
  \cellcolor[HTML]{00009B}{\color[HTML]{FFFFFF} 0.87} &
  \cellcolor[HTML]{00009B}{\color[HTML]{FFFFFF} 0.80} &
  \cellcolor[HTML]{00009B}{\color[HTML]{FFFFFF} 1.5} & 
  \cellcolor[HTML]{00009B}{\color[HTML]{FFFFFF} 9}\\ \cline{1-2}
jaccard &
  jaccard &
  \cellcolor[HTML]{00D2CB}{\color[HTML]{000000} 0.63} &
  \cellcolor[HTML]{00D2CB}{\color[HTML]{000000} 0.75} &
  \cellcolor[HTML]{00D2CB}{\color[HTML]{000000} 0.70} &
  \_  &
  \_\\ \hline
\end{tabular}%
}
\caption{Offline evaluation on search relevance agreement and the Diff and Full Batch inference latencies in hours. Darker colors represent more favorable scores.}
\label{tab:results1}
\end{table}

From the results in Table \ref{tab:results1} we can infer that of the bi-encoders the most successful one is the contrastive loss trained bi-encoder. This beats the IRNS and the Softmax trained bi-encoders by a significant margin in the F1 scores. For the bi-encoders of IRNS and Constrative we see a very high recall. For cross-encoders we see that the best performance is achieved by \texttt{bert-mini}, beating all the models in their precision and F1 scores, with a very high recall as well. Our smallest model \texttt{bert-tiny} has the 2nd highest precision and F1 score of 0.74 and 0.80 respectively. This shows that even the smallest cross-encoders can beat the relatively more complex bi-encoders.  We also see that Contrastive Loss is a better option in training the bi-encoders for the relevance computation than Softmax and IRNS.

Bi-encoders and cross-encoders  have different serving contracts --- bi-encoders perform inference for item and keyphrases independently while cross-encoders work on the combination of the two. Our production architecture illustrated in Figure \ref{fig:prod_arch} features two components --- \textit{Near Real-Time (NRT)} Inference and Batch Inference. Batch inference primarily serves items with a delay, whereas NRT serves items on an urgent basis, such as items newly created or revised by sellers. The batch inference is done in two parts: 1) for all items and keyphrases, and 2) daily differential (Diff), i.e. the difference of all new items and keyphrases created/revised and then merged with the existing ones. The NRT inference for our relevance model is written into the service code using DeepJava \cite{Deepjavalibrary} and onnx serving. The NRT service is triggered by the event of creation or revision, behind a Flink processing window and feature enrichment.  

The full batch inference is around 2.5 billion items, 1 billion keyphrases and around 100 billion item-keyphrase pairs, while for daily differential we serve around 20 million items, 3 million keyphrases and around 7 billion item-keyphrase pairs. Since full batch inference has to be carried out once, the Diff inference latency is what decides if we can put a model in production (we still provide the numbers in Table \ref{tab:results1}). The Diff inference latency for bi-encoders is 1 hour while for \texttt{bert-tiny} and \texttt{bert-mini} it is 1.5 and 2.5 hours, respectively. All latency numbers are benchmarked using PySpark \cite{spark} (1000 executors, 20g memory and 4 cores) with transformers \cite{wolf-etal-2020-transformers} and onnxruntime \cite{onnxruntime} libraries. While GPU options were considered, they proved to be unsustainable due to their cost. Taking into account our serving requirements, the \texttt{bert-tiny} model, which has a overall decent performance, was chosen for release. In terms of latency, it matches the bi-encoders and offers quality close to that of the top-performing \texttt{bert-mini} model.

The \texttt{bert-tiny} model was released into production following an A/B test against the Jaccard model in production. The \texttt{bert-tiny} model delivered on all fronts with impressions, clicks, bought items, CVR or conversion rate (bought items / click) increasing by 20.17\%, 13.01\%, 26.04\%, 11.6\%, respectively (all with p-value less than 0.05) while CTR or click-through rate (clicks / impressions) remained neutral (p-value=0.17). Overall, we reduced the number of keyphrases surfaced to sellers by 7\% and increased the efficacy of our recommendations, as is evident in these numbers. A follow-up analysis showed that the \texttt{bert-tiny} model decreased the False Positives (i.e., Search irrelevant keyphrases that passed the Advertising filter) and also the False Negatives (Search relevant that were filtered out) while increasing the True Positives (Search relevant keyphrases that pass the Advertising filter). Of the False Negatives that were reduced by \texttt{bert-tiny}, a large portion of them were shorter head keyphrases, which are very desirable for advertisers due to their targeting potential.

\section{Conclusions}

In this study, we discuss the limitations of using clicks solely as a training signal and introduce the concept of \textit{middleman bias} in the context of relevance of advertiser keyphrase recommendations. We reinterpret this task as a complex interaction between two dynamical systems --- Advertising, which suggests keyphrases; and Search, which manages the items entering the auction for these keyphrases.  Our findings indicate an imbalance in this interaction, with Search ultimately determining relevance of items to keyphrases, thus underscoring the importance of alignment with Search. This research also offers a comparative study of two well-known encoder-based frameworks in modeling this alignment: bi-encoders and cross-encoders of varying sizes, training objectives, and losses. We show that even very simple 2-layer cross-encoders significantly outperform more intricate bi-encoders in relevance modeling. Additionally, we highlight that while bi-encoders are considerably faster, utilizing lighter cross-encoders can narrow the latency difference and provide better outcomes.

\bibliographystyle{ACM-Reference-Format}
\balance
\bibliography{refs}
\end{document}